# Quantitative pulsatility measurements using 3D Dynamic Ultrasound Localization Microscopy

Chloé Bourquin[1], Jonathan Porée[1], Brice Rauby[1], Vincent Perrot[1], Nin Ghigo[1], Hatim Belgharbi[1,2], Samuel Bélanger[3], Gerardo Ramos-Palacios[4], Nelson Cortés[5], Hugo Ladret[5,6], Lamyae Ikan[5], Christian Casanova[5], Frédéric Lesage[7,8], and Jean Provost[1,8]

[1]Department of Engineering Physics, Polytechnique Montréal, Montréal, QC H3T 1J4, Canada
[2]Department of Biomedical Engineering, University of North Carolina, Chapel Hill, NC 27599, USA
[3]Labeo Technologies Inc., Montréal, QC H3V 1A2, Canada
[4]Montreal Neurological Institute, McGill University, Montréal, QC H3A 2B4, Canada
[5]School of Optometry, University of Montreal, Montréal, QC H3T 1P1, Canada
[6]Institut de Neurosciences de la Timone, UMR 7289, CNRS and Aix-Marseille Université, Marseille, 13005, France
[7]Department of Electrical Engineering, Polytechnique Montréal, Montréal, QC H3T 1J4, Canada
[8]Montreal Heart Institute, Montréal, QC H1T 1C8, Canada

## Abstract

A rise in blood flow velocity variations (i.e., pulsatility) in the brain, caused by the stiffening of upstream arteries, is associated with cognitive impairment and neurodegenerative diseases. The study of this phenomenon requires brain-wide pulsatility measurements, with high penetration depth and high spatiotemporal resolution. The development of Dynamic Ultrasound Localization Microscopy (DULM), based on ULM, has enabled pulsatility measurements in the rodent brain in 2D. However, 2D imaging accesses only one slice of the brain and measures biased velocities due to 2D projection. Herein, we present 3D DULM, which can extract quantitative pulsatility measurements in the cat brain with craniotomy and in the whole mouse brain through the skull, showing a wide range of flow hemodynamics in both large and small vessels. We highlighted a decrease in pulsatility along the vascular tree in the cat brain, and performed an intra-animal validation of the method by showing consistent measurements between the two sides of the Willis circle in the mouse brain. Our study provides the first step towards a new biomarker that would allow the detection of dynamic abnormalities in microvessels in the whole brain volume, which could be linked to early signs of neurodegenerative diseases

## 1. Introduction

Arterial stiffening and the subsequent rise in pulsatility in downstream microvessels is known to be associated with neurodegenerative diseases such as Alzheimer's (AD) [1]. AD patients were not only found to have a higher pulsatility index (PI) [2] in cerebral arteries compared to non-demented subjects [3], but a rise in pulsatility in non-demented subjects was also linked to a significant cognitive decline a few years later [4]. Thus, assessing the blood flow pulsatility in main cerebral arteries with Transcranial Doppler (TCD) ultrasound was proposed as a tool to monitor the progression of neurodegenerative diseases such as AD [5]. However, even at high frequency [6], TCD only allows for pulsatility measurements in major vessels, which may not be sufficient to report on pulse transmission to downstream microvessels and potential damage. Other brain imaging techniques such as optical coherence tomography [7] or two-photon microscopy [8] can also perform pulsatility measurements in microvessels at the surface of the brain, but do not provide enough penetration depth to achieve brain-wide pulsatility measurements.

We recently proposed a novel method, called Dynamic Ultrasound Localization Microscopy (DULM) [9], that could address this issue. Based on Ultrasound Localization Microscopy (ULM) [10]–[12], it consists in injecting microbubbles in the blood stream to follow them in the vascular tree using an ultrafast ultrasound system. After a few minutes of acquisitions, the microbubbles are localized and tracked across frames to produce a super-resolved map of the vasculature beyond the acoustic diffraction limit, which usually restricts the imaging resolution in conventional echography. The main difference between ULM and DULM is DULM's temporal resolution, enabled by the triggering of the ultrasound system on the electrocardiogram (ECG), which allows us to know the position of each frame within the cardiac cycle.

ULM can map, in 3D, the angioarchitecture of the whole rodent brain [13]–[15]. It was also successfully tested in the human brain in 2D [16], to retrieve dynamic biomarkers such as systolic and diastolic velocities. Using DULM, brain-wide

This work was supported in part by the Institute for Data Valorization (IVADO), in part by the Canada Foundation for Innovation under Grant 38095, in part by the Canadian Institutes of Health Research (CIHR) under Grant 452530, in part by the New Frontiers in Research Fund under Grant NFRFE-2018-01312 and in part by the Natural Sciences and Engineering Research Council of Canada (NSERC) under Grant RGPIN-2019-04982. The work of C. Bourquin, J. Porée, B. Rauby, V. Perrot and H. Belgharbi was supported in part by IVADO, in part by the TransMedTech Institute, in part by the Fonds de recherche du Québec — Nature et technologies, in part by the Quebec Bio-Imaging Network, and in part by the Canada First Research Excellence Fund (Apogée/CFREF). This research was enabled in part by support provided by Calcul Québec (calculquebec.ca) and the Digital Research Alliance of Canada (alliancecan.ca). (Corresponding author: Jean Provost).

pulsatility measurements in microvessels in 2D were also shown to be possible in vivo in multiple instances, e.g., in a rat brain with craniotomy and a mouse brain through skull and skin [9], and in the rat heart after motion correction [17]. More recently, at another temporal scale, ULM could measure changes in the blood flow dynamics due to functional activation in a rat brain with craniotomy in 2D [18]. However, a 2D image does not provide an overview of the brain-wide pulsatility in the vascular tree, from major arteries to microvessels.

Herein, we show that 3D DULM can perform quantitative pulsatility measurements in the cat brain with craniotomy and in the whole mouse brain through the skull. In the cat brain, we showed a significant pulsatility attenuation in the vascular tree, from large feeding vessels to downstream smaller ones. An intra-animal validation of the measurements was performed in the mouse brain using the symmetry of the Willis circle. We demonstrate that 3D DULM can produce highly resolved dynamic maps of the blood flow and provides a novel method to better understand its dynamics in the vascular tree, by performing deep, non-invasive pulsatility measurements in the whole brain.

## 2. Methods

### 2.1 In vivo experiments

#### 2.1.1 Ethics

All surgical and experimental procedures were undertaken according to the guidelines of the Canadian Council on Animal Care and were approved by the Ethics Committee of the University of Montreal (CDEA 19-008 and 19-064).

#### 2.1.2 Animals' preparation

Cat: detailed procedures are described in [19]. Before surgery, the cat received a solution of atropine (0.1 mg/kg) and acepromazine subcutaneously. Anesthesia was induced with 3.5% isoflurane in a 50:50 gas mixture of $O_2$ and $N_2O$. A catheter was placed in the cephalic vein to provide intravenous access. A tracheotomy was performed prior to the transfer of the animal to the stereotaxic apparatus. Following anesthetic induction, isoflurane concentration was 1.5% during surgical procedures. During recording sessions, the anesthesia was changed to halothane (0.5–0.8%) in a 30:70 gas mixture of $O_2$ and $N_2O$. An intravenous bolus injection of 2% gallamine triethiodide was administered to induce muscular paralysis, and subsequently, the animal was placed under artificial ventilation. A 1:1 solution of 2% gallamine triethiodide (10 mg/kg/h) in 5% of dextrose in lactated ringer was continuously administered intravenously. Expired levels of $CO_2$ were maintained between 35 and 40 mmHg by adjusting the tidal volume and respiratory rate. Temperature, $SpO_2$ and heart rate were monitored during the whole experiment. 3D in vivo DULM imaging of the cat brain was performed after a three-day optical imaging experiment, during which the animal was immobilized and under anesthesia (data not shown for the current study). Just before the ultrasound imaging, a 15 × 15 mm² craniotomy window was performed on the contralateral side of the one used for optical imaging procedures. A durotomy was done to access the brain surface, where ultrasound gel was applied. The animal's heart rate was monitored throughout the procedure (Labeo Technologies Inc., QC, Canada). A 2-mL solution of microbubbles ($1.2 \times 10^{10}$ microbubbles per milliliter, Definity, Lantheus Medical Imaging, MA, USA) diluted in saline (1:1) was injected as a bolus in the paw using a catheter.

Mouse: After sedation with chlorprothixene (5 mg/kg), the mouse was anesthetized with urethane (1.0–1.5 g/kg, i.p., at 10% w/v in saline). Atropine (0.05 mg/kg) was also injected to reduce secretions in the airway. After a tracheotomy was performed, the animal was placed on a platform (Labeo Technologies Inc., QC, Canada), where temperature and heart rate were monitored throughout the experiment. Before exposing the skull, a subcutaneous injection of lidocaine 2% was applied. The skin above the skull was removed, where ultrasound gel was applied. A 50-µL solution of microbubbles ($1.2\times10^{10}$ microbubbles per milliliter, Definity, Lantheus Medical Imaging, MA, USA) diluted in saline (1:1) was injected as a bolus in the tail vein.

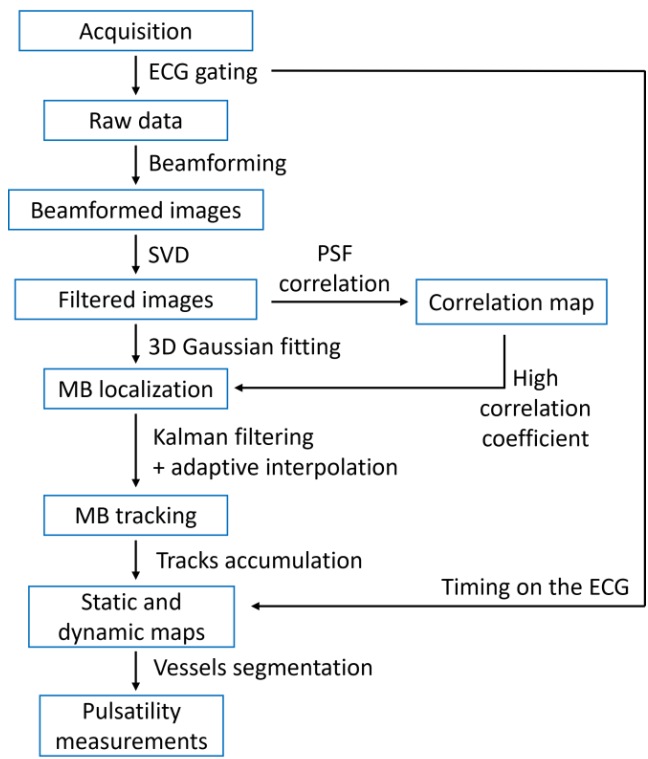

Figure 1. 3D DULM pipeline (see Methods). SVD: Singular Value Decomposition filtering, PSF: Point Spread Function, MB: microbubbles.

Table 1. 3D ultrasound imaging sequence gated on the ECG

| Sequence parameters | Cat brain | Mouse brain |
| --- | --- | --- |
| Probe | 8-MHz 2D matrix probe | |
| # of groups acquired | 800 | |
| # of transmits / receives | 4 per frame at angle 0° | |
| # of cycles | 4 | 2 |
| Frame rate (Hz) | 1000 | 2000 |
| Penetration depth (mm) | 37 | 19 |
| # of frames acquired per group | 600 | 400 |
| # of cardiac cycles per group | 1.6 | 1.7 |
| Total acquisition time (min) | 15 | 17 |
| Effective acquisition time (min) | 8 | 3 |

### 2.1.3 Ultrasound Acquisitions

3D in vivo ultrasound imaging of both species (see Table 1) was performed using an 8-MHz 2D matrix probe (Verasonics, WA, USA), with $32 \times 32$ elements. This probe was connected to a programmable ultrafast ultrasound system (Vantage 256, Verasonics, WA, USA). Each frame was completed with a succession of four Transmits/Receives: for one Transmit/Receive, we sent plane waves at angle 0° with all 1024 elements of the probe and received them with a sub-aperture of 256 elements (¼ of the probe). Such a sequence was primarily designed to image the cat brain and keep a high frame rate (1000 Hz) despite a high penetration depth (37 mm). The sequence was then adapted for the mouse brain, in which the 19-mm depth allowed for a frame rate of 2000 Hz.

Both animals were imaged using an ultrasound sequence synchronized with the ECG. The frames were acquired in groups lasting less than one second for ~ 15 min, with pauses for data transfer and saving between each group. Without taking the pauses into account, the effective acquisition time was about 3 min for the mouse brain and 8 min for the cat brain. The first frame of each group started when the monitoring platform (Labeo Technologies Inc., QC, Canada) detected an R-wave in the ECG. Such a gated sequence allowed us to record, for each group, 1.6 cardiac cycles in the cat and 1.7 cardiac cycles in the mouse.

## 2.2 Data processing

### 2.2.1 DULM Processing

As in [9], after beamforming, the noise was first normalized by dividing each pixel by the square root of the number of channels contributing to each pixel [20]. We then used a Singular Value Decomposition (SVD) filter to remove tissue by setting the first 20 singular values to zero. To enhance the microbubbles' signal, the images were correlated with the Point Spread Function (PSF) of the system, which was simulated using an in-house GPU implementation of the SIMUS simulation software [21]. The microbubbles were localized using a Gaussian least square fitting. 1024 microbubbles were detected in each frame, and only microbubbles with a high correlation coefficient were kept for the following tracking algorithm. All these pre-processing algorithms (beamforming, tissue cancellation and localization, see Fig. 1) were implemented on MATLAB 2021a (The MathWorks, Inc., Natick, MA).

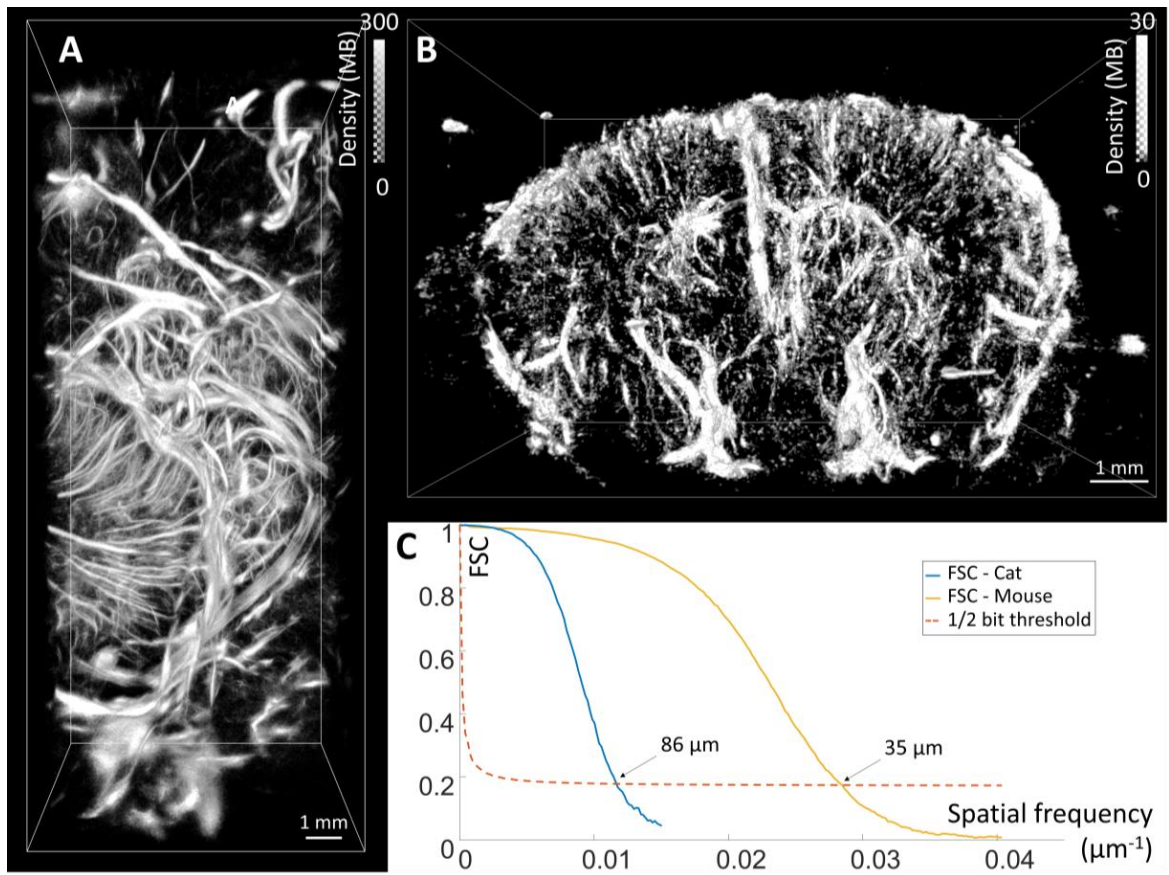

Figure 2. Resolution measurements based on the Fourier Shell Correlation (FSC). A. Static density map of the cat brain. B. Static density map of the mouse brain. C. FSC calculation. The measured resolutions are 86 µm in the cat brain and 35 µm in the mouse brain.

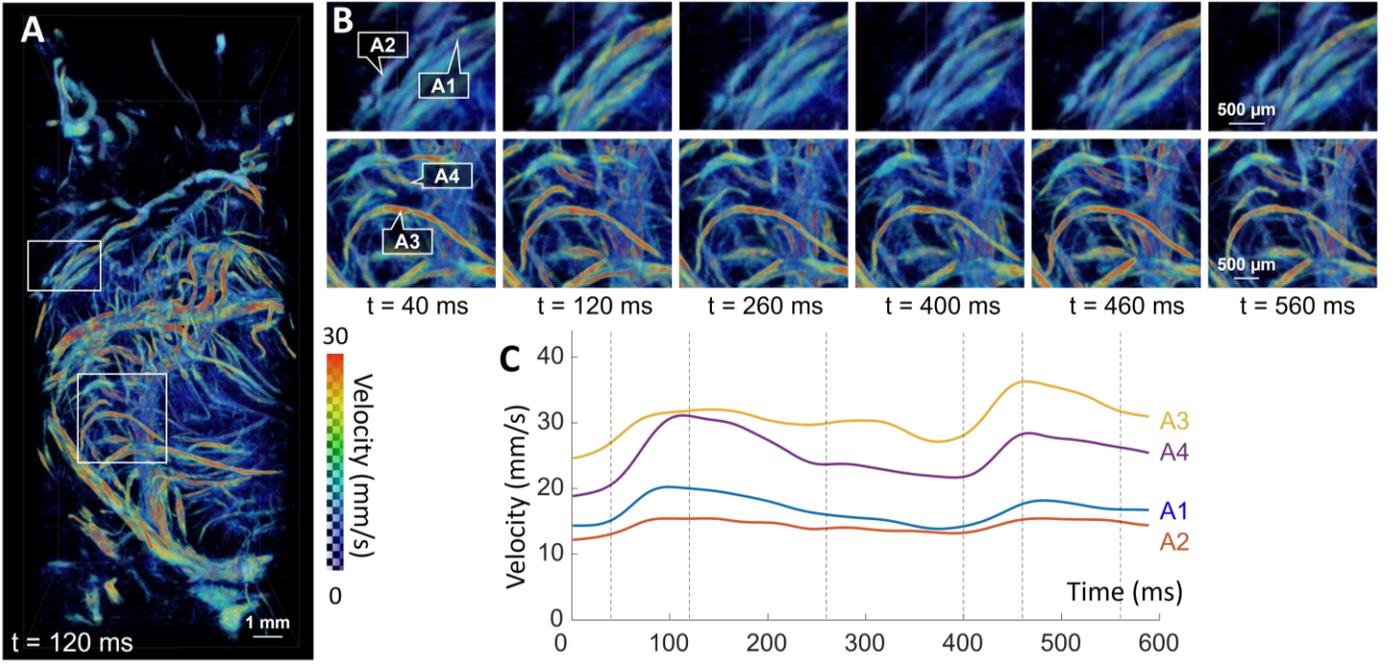

Figure 3. Pulsatility measurements performed in the 3D volume of the cat brain. A. Projection of the 3D velocity map of the cat brain, taken at t = 120 ms during the acquisition. Two regions of interest (ROI) were selected in the brain. B. Qualitative velocity variations observed in selected vessels (A1-4) observed in the ROI selected in A, at different time points during the acquisition. C. Velocity variations over time (600 ms) of vessels A1-4. Vertical lines correspond to the timing of the images displayed in B.

A Kalman filter was then used to track the microbubbles. The algorithm was inspired by the one described in [22] and adapted to 3D on Python 3.9. For each new microbubble, we allocated a new Kalman filter to predict its position and velocity in the next frame. The prediction was then updated in the following frames, depending on the previous positions and velocities. As mentioned in [22], we then used an adaptive interpolation to bridge the gaps in the tracks between distant microbubbles: the interpolation factor was set to get, in each track, a maximum distance between two consecutive microbubbles of ~ 2 µm in the mouse brain, and ~ 5 µm in the cat brain. This approach helped us to recover microbubbles positions, even in vessels where only a few microbubbles were detected, and to follow microbubbles with different behaviors (fast, slow, accelerating, etc.) from one frame to another, to finally calculate their velocity. We did not add constraints on the microbubbles' trajectory or acceleration.

### 2.2.2 Dynamic maps

As described in [9], after tracking, the microbubbles' positions and velocities were averaged across all the groups according to their timing in each ECG-gated acquisition. By doing so, we obtained dynamic maps of the microbubbles flowing in the blood stream, lasting more than one cardiac cycle. Dynamic maps were further processed as videos using the Amira 3D software 2021.2 (Thermo Fisher) (e.g., Supplementary videos 1 and 2).

### 2.2.3 Resolution calculation

For analysis, the microbubbles' positions were accumulated to compute a static density map (Fig. 2 (A) and (B)), corresponding to the sum of all the microbubbles detected in each pixel during the whole acquisition. These static maps were used for resolution estimation and vessel segmentation, which was used for the quantitative pulsatility analysis (see below).

The 3D spatial resolution was calculated using the Fourier Shell Correlation (FSC), also known in 2D as the Fourier Ring Correlation (FRC), which is a tool used to estimate spatial resolution in Single Molecule Localization Microscopy. It was recently introduced to calculate the resolution in ULM in 2D [23], and we adapted it in 3D from https://github.com/bionanoimaging/cellSTORM-MATLAB/ [24].

For both species, we randomly separated the tracks across all the buffers into two datasets $Im_1$ and $Im_2$, to obtain two independent reconstructions of the same brain volume. We then calculated the FSC (Fig. 2 (C)) as the normalized correlation of their spectrum $F_1$ and $F_2$ along all the voxels $r_i$ located at the radius $r$:

$$FSC(r) = \frac{\sum_{r_i \in r} F_1(r_i) \cdot F_2(r_i)^*}{\sqrt{\sum_{r_i \in r} |F_1(r_i)|^2 \cdot \sum_{r_i \in r} |F_2(r_i)|^2}} \quad (1)$$

Table 2. Examples of Pulsatility Index measurements performed by 3D DULM in the cat brain

| Vessel | Diameter ⌀ (µm) | Mean velocity (mm/s) | Pulsatility Index |
|---|---|---|---|
| A1 | 212 | 16.81 | 0.30 |
| A2 | 96 | 14.35 | 0.19 |
| A3 | 154 | 30.80 | 0.27 |
| A4 | 156 | 25.41 | 0.37 |

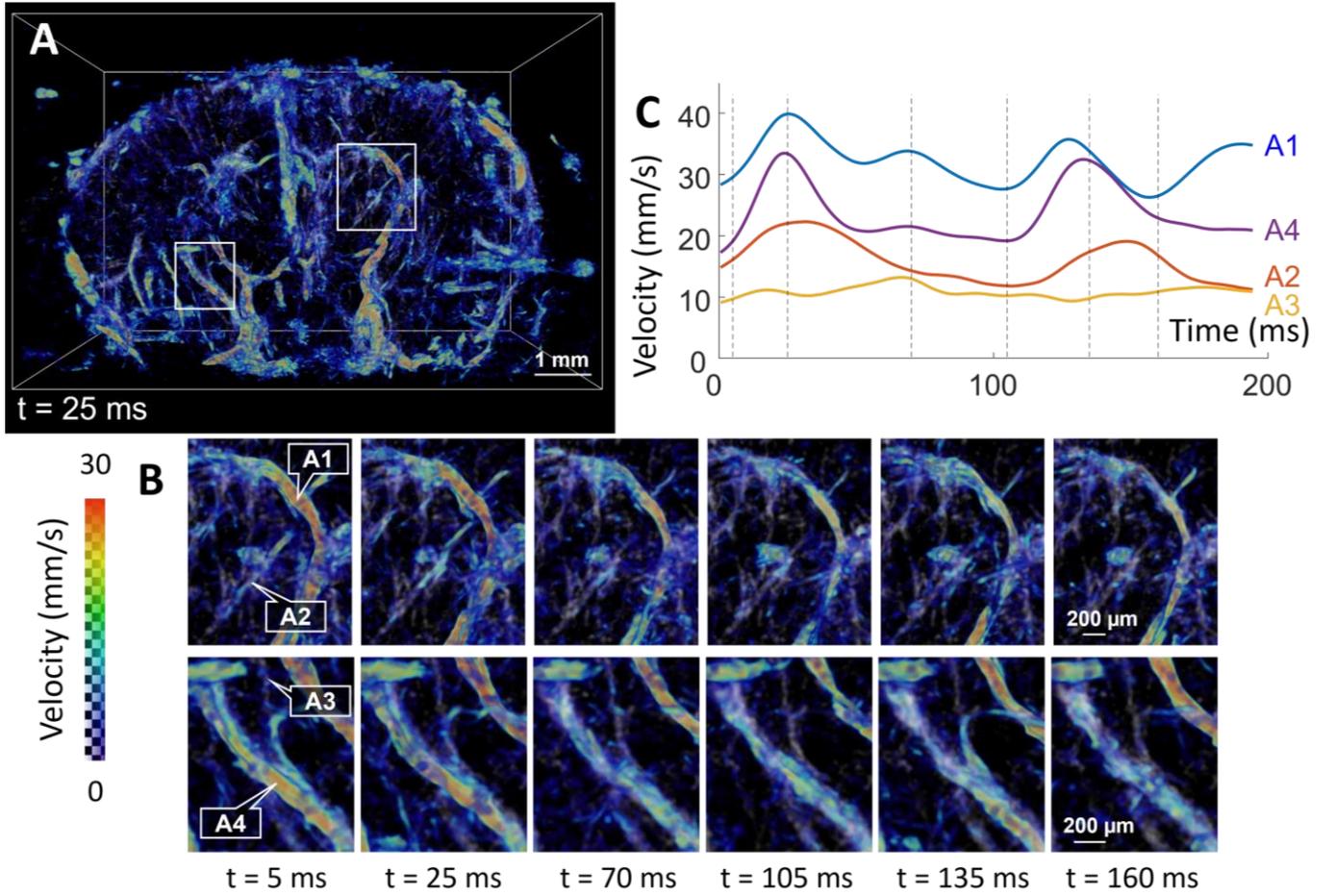

Figure 4. Pulsatility measurements performed in the 3D volume of the mouse brain. A. Projection of the 3D velocity map of the mouse brain, taken at t = 25 ms during the acquisition. Two regions of interest (ROI) were selected in the brain. B. Qualitative velocity variations observed in selected vessels (A1-4) observed in the ROI selected in A, at different time points during the acquisition. C. Velocity variations over time (200 ms) of vessels A1-4. Vertical lines correspond to the timing of the ROI displayed in B.

The resolution was estimated as the intersection between the FSC curve and the 1/2-bit-threshold curve, as a function of $r$. This value was compared with the axial and lateral resolutions in conventional echography, which are restricted by the diffraction limit, and more particularly by the wavelength $\lambda$ ($\lambda \sim 200$ μm at 8 MHz).

On the one hand, the axial resolution is approximately equal to half of the spatial pulse length, i.e $\lambda$ in the mouse brain and $2\lambda$ in the cat brain (see Table 1). On the other hand, the lateral resolution $LR$ is given by:

$$LR = \lambda f / d \quad (2)$$

where $d$ is the aperture and $f$ the focal distance.

For $d = 48\lambda$, as used in this study, the lateral resolution at half the total imaging depth (~ 20 mm in the cat brain and ~ 10 mm in the mouse brain) was thus approximately equal to the axial resolution: 400 μm (i.e., $2\lambda$) in the cat brain and 200 μm (i.e., $\lambda$) in the mouse brain.

### 2.2.4 Segmentation and pulsatility measurements

Vessels were segmented in the static maps using a Hessian filter. High frequencies in the velocity signal were then filtered out (moving average of 100 values in the cat brain, or 75 values in the mouse brain, which corresponds to a frequency cutoff of 10 Hz and 26.67 Hz respectively). The resulting filtered signal was sufficient to sample the cardiac cycles of both species, which had a frequency of 2.67 Hz for the cat and 8.4 Hz for the mouse.

The pulsatility index (PI) [2] was computed as the difference between the peak systolic velocity (PSV) and the end-diastolic velocity (EDV), divided by the mean flow velocity (MFV):

$$PI = \frac{PSV - EDV}{MFV} \quad (3)$$

Table 3. Examples of Pulsatility Index measurements performed by 3D DULM in the mouse brain

| Vessel | Diameter ⌀ (μm) | Mean velocity (mm/s) | Pulsatility Index |
|---|---|---|---|
| A1 | 91 | 32.13 | 0.31 |
| A2 | 46 | 15.95 | 0.46 |
| A3 | 38 | 10.84 | 0.24 |
| A4 | 113 | 23.81 | 0.58 |

The PI reflects the variations of the blood flow velocity compared to its mean velocity. The pulsatility can also be qualitatively characterized by the velocity waveform itself, which presents a pulsatile pattern synchronized with the cardiac rhythm for a high pulsatility [25].

*2.2.5    Willis circle analysis*

In the mouse brain static map (Fig. 6 (A)), the shape of the Willis circle structure could be recognized. This symmetrical structure corresponds to the crossing of several main arteries of the brain, in the left and the right hemispheres: the posterior communicating artery (PComA), the internal carotid artery (ICA), the anterior cerebral artery (ACA) and the middle cerebral artery (MCA). These arteries were segmented as described above using a Hessian filter, to compute their velocity variations and pulsatility indices.

## 3. Results

*3.1    Resolution calculation*

Static density maps obtained after processing can be observed in Fig. 2 (A) and (B).

After calculating the FSC for both species, the resolution in these maps was estimated at 86 µm in the cat brain and 35 µm in the mouse brain (Fig. 2 (C)), which correspond to $\lambda/2$ and $\lambda/6$ respectively.

*3.2    Whole-brain qualitative pulsatility measurements*

Dynamic maps generated by 3D DULM (Supplementary videos 1 and 2) show microbubbles flowing in the bloodstream in the whole brain in 3D and their velocity variations during the cardiac cycle. The pulsatility can be qualitatively observed in the videos during the cardiac cycle in some large and small vessels shown by the arrows, in both animals.

Figures 3 (A) and 4 (A) show an example of the 3D velocity map obtained at t = 120 ms in the cat brain and t = 25 ms in the mouse brain. Two regions of interest (ROI) were extracted from these maps and displayed in Figures 3 (B) and 4 (B) at different time points of the cardiac cycles. Velocity variations could be seen in some large and small vessels.

*3.3    Whole-brain quantitative pulsatility measurements*

A few vessels with different diameters and mean velocities were segmented from these ROI, in the cat and the mouse brains, and their velocity variation through time was calculated (Fig. 3 (C) and 4 (C)). By doing so, we obtained an overview of the performance of the method applied to different flow patterns: large and small vessels, and fast and slow flow. Blood flow pulsatility was quantified by calculating the PI [2] (see Methods). Examples of quantitative pulsatility measurements that could be performed in these vessels are shown in Tables 2 and 3.

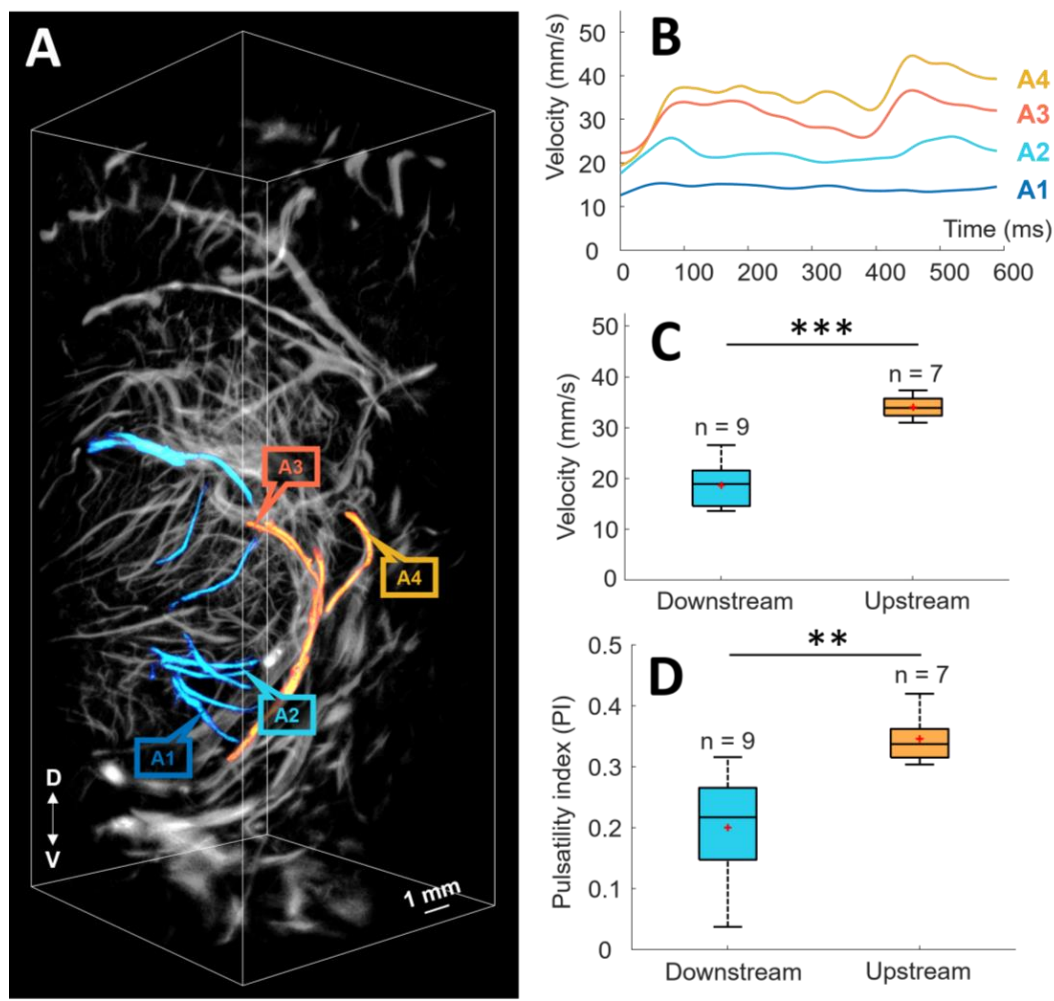

Figure 5. Pulsatility measurements performed in the cat brain. A: Static density mapping with vessel labels. A3 and A4 correspond to upstream feeding vessels, and A1 and A2 correspond to downstream vessels. B: Velocity variations over time (600 ms) in vessels. C. Comparison of mean velocities between upstream feeding vessels (orange) and downstream vessels (blue) (P < 0.001, Student t-test). D. Comparison of PI between upstream feeding vessels (orange) and downstream vessels (blue) (P < 0.01, Student t-test).

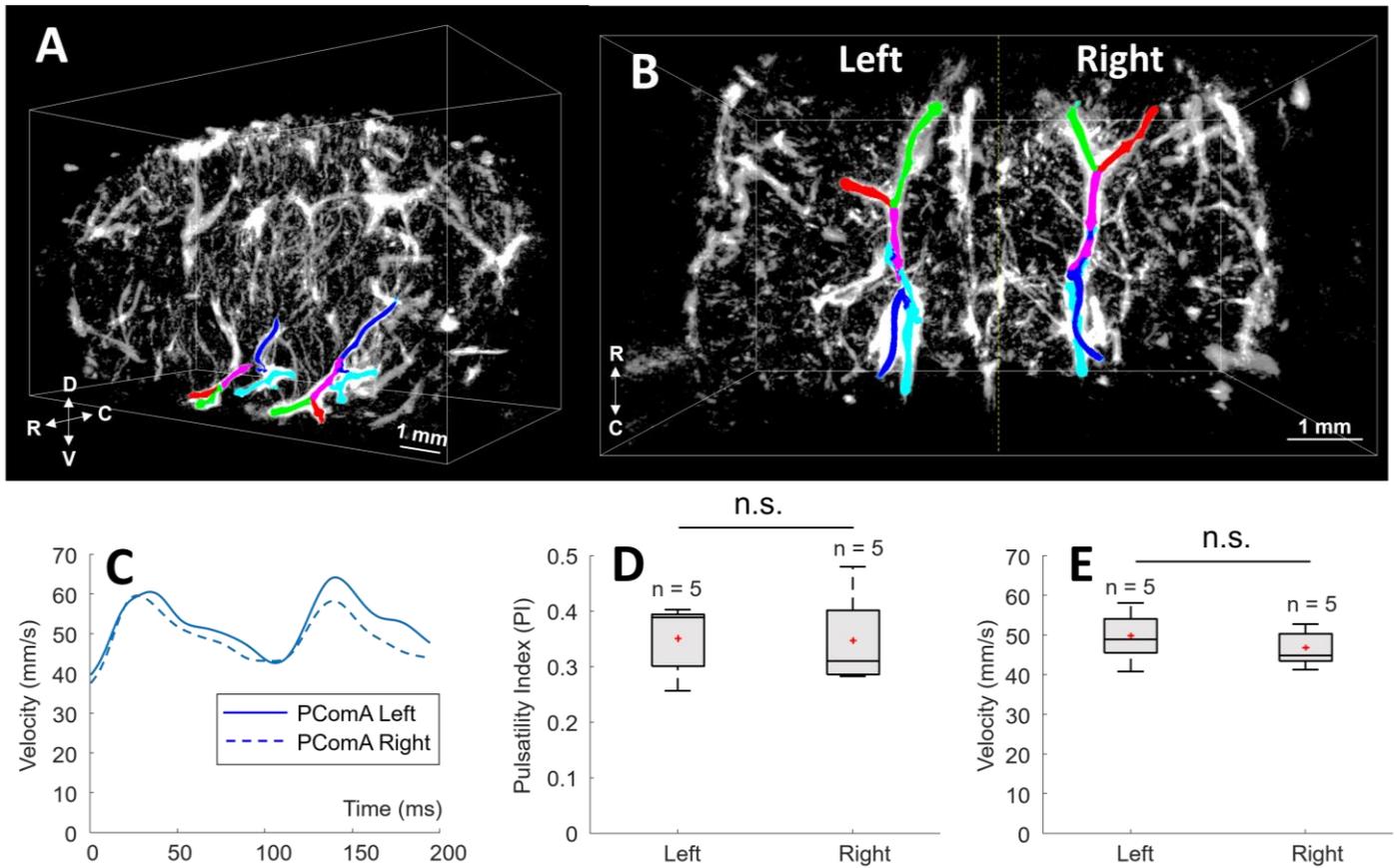

Figure 6. Pulsatility measurements in the Willis circle. A/B. Willis circle identification and segmentation. C. Velocity variations in time (200 ms) for the PComA identified in both sides the Willis circle. D. Mean velocities measured in the vessels (n = 5) identified in both sides of the Willis circle (Student t-test, P = 0.93). E. Pulsatility indices calculated in the vessels (n = 5) identified in both sides of the Willis circle (Student t-test, P = 0.42).

A reproducible and synchronized pulsatility pattern of about one and a half cardiac cycles was observed in both species, even in microvessels with a diameter close to the resolution provided by the FSC calculation (A2: ⌀ 46 µm in the mouse, A2: ⌀ 96 µm in the cat). Some microvessels also presented a flat velocity variations pattern, with a low PI, such as A3 (⌀ 38 µm, PI: 0.24) in the mouse brain.

In some vessels, the presence of a dicrotic notch was observed, corresponding to a decrease in arterial pressure occurring immediately after the systole. In the cat brain, this dicrotic notch could be seen in vessels A3 and A4; in the mouse brain, it could be observed in vessels A1 and A4.

*3.4 Pulsatility and velocity attenuations along the vascular tree*

Large upstream feeding vessels and their subsequent downstream vessels were segmented in the 3D volume of the cat brain (Fig. 5 (A)). Pulsatility measurements were performed in both (n = 7 and n = 9, respectively). Examples of the velocity variations in time obtained in these vessels can be observed in Fig. 5 (B).

In Fig. 5 (B), in upstream vessels (A3 and A4), a strong, synchronized, and consistent pulsatility pattern can be observed in the velocity variations. In downstream vessel A2, a smaller but still noticeable pulsatility pattern can be seen. In A1, no pulsatility pattern can be observed.

Mean velocity and PI were calculated for both groups of vessels (upstream and downstream) in Fig. 5 (C, D). A significant difference between the two groups, in both, mean velocity (P < 0.001) and PI (P < 0.01), was observed.

*3.5 Intra-animal validation in the Willis circle*

Several main arteries belonging to the symmetrical structure of the Willis circle were recognized in the mouse brain and were segmented (Fig. 6 (A, B)). We found similar velocity waveforms for each vessel between the two sides (left and right) of the Willis circle, with a high pulsatility index and a high velocity, as shown for example in Fig. 6 (C) for the PComA.

The pulsatility and mean velocity measurements between the two sides of the brain were compared (Fig. 6 (D)): the PI of all the left vessels (1) was found to be 0.35 ± 0.06, and 0.35 ± 0.08 for the right vessels (2); the mean velocity was found to be 50 ± 6 mm/s for the left (1) and 47 ± 5 mm/s for the right (2). A Student t-test was performed, and the differences in PI and mean velocity between the left and the right sides of the Willis circle were found to be non-significant (resp. P = 0.93 and P = 0.42).

## 4. Discussion

In this study, 3D Dynamic Ultrasound Localization Microscopy (3D DULM) was used to image the blood velocity as a function of time and to measure pulsatility indices in a cat brain and a mouse brain, with and without craniotomy, respectively.

More specifically, with 3D DULM, not only highly resolved maps of the brain vasculature were produced (see Fig. 2), but also cine-loops of flowing microbubbles in the entire volume of the brain, reflecting the blood flow variations during the cardiac cycle. An example of these cine-loops was provided in Supplementary videos 1 and 2, in both animals.

This qualitative observation was confirmed with quantitative measurements: in Fig. 3 and Fig. 4, velocity variations over time and PI calculations were extracted for different vessels in the whole brain, and a pulsatile pattern synchronized with the cardiac rhythm of the animals was recognized in the velocity waveforms, even for slow flows and small vessels.

In the cat brain (Fig. 5), we observed a significant attenuation of the mean velocity and the pulsatility index along the vascular tree, between upstream feeding vessels and downstream subsequent ones.

In Fig. 6, the main arteries in the mouse Willis circle were identified. Their mean velocities and PI were measured and found to be high and reproducible between the left and the right sides of the brain, which is consistent with the literature [26], [27].

3D DULM could provide both qualitative and quantitative information on the blood flow pulsatility in the brain, with 1) dynamic 3D maps of microbubbles circulating in the whole brain, reflecting changes in blood flow during the cardiac cycle; 2) the measurement of velocity variations over time in any observable vessel in the vascular tree, with a 10-ms temporal resolution and a sub-wavelength spatial resolution (< 100 µm, i.e. $\lambda/2$); and 3) the PI estimation in these vessels, to quantify the pulsatility. Until now, in small animals, the pulsatility could only be quantified in large vessels [28], in small vessels at the surface of the brain [7], [8], or in a 2D section of the brain using 2D DULM [9]. With 3D DULM, we were able to extract pulsatility measurements all along the vascular tree, in depth, in the whole brain in 3D, with a high spatiotemporal resolution. With this technique, dynamic measurements could be extracted in a range of vessels that were inaccessible for non-invasive imaging modalities before. Such a modality could be used for example to study the pulsatility propagation in the brain vasculature and to better understand its impact on microvessels.

The study presented herein is however not without limitations. Our experimental set up relied on bolus injections, which leads to a variation in the microbubble concentration over time during the acquisition. Although bolus injection is a frequently used method in ULM, a continuous infusion could have been used [12], which may be optimized to improve resolution and decrease the acquisition time by better controlling microbubble concentration [29].

The maximal PRF we could achieve was limited by multiplexing: each frame demanded to perform four Transmits/Receives at angle 0°. A combination of several ultrasound systems, such as the one described in [30], may overcome such a limitation and allow for performing several angles while keeping a high frame rate and a large penetration depth.

The ~ 15-min sequences also included pauses for the transfer and the saving of the data. However, as it is mentioned in [13], this drawback of performing volumetric images will decrease in the next years with the increase of transfer speed and computational power: for example, without the pauses in the sequence, the mouse brain could have been imaged with a 3-min acquisition, dividing the acquisition time five-fold.

The achieved temporal resolution of ~ 0.1 second in this study was not sufficient to measure an increase in the time of arrival of the pulsatile flow along the vascular tree, but this could be another interesting biomarker to measure in the study of neurodegenerative diseases.

Moreover, the sub-wavelength spatial resolutions we obtained (86 µm in the cat brain and 35 µm in the mouse brain, corresponding to $\lambda/2$ and $\lambda/6$ respectively) are not as small as what can be achieved in usual ULM studies in 2D. This can be explained by different factors. First, only one angle (at 0°) was used in the sequences to keep a high frame rate despite a large penetration depth in the cat brain. Hence, the quality of the images and the resolution could have been improved by using several compounding angles, in the lateral and the elevation directions. Also, no aberration or motion correction algorithms were used, which could have improved image quality [16], [31]. Additionally, compared to the diffraction-limited axial and lateral resolutions that can be obtained (see Methods) with the $32 \times 32$ matrix probe that we used, the resolution that we achieved shows a gain of a factor ~ 4 in the cat brain and a gain of a factor ~ 6 in the mouse brain. These resolutions are comparable with the ones obtained with a similar probe and a similar sequence in [32], which achieved a 52-µm resolution at an 11-mm depth in a chicken embryo brain, corresponding to a gain of a factor ~ 4, in both lateral and axial directions.

## 5. Conclusion

3D DULM is a new ultrasound imaging technique that generates dynamic maps of flowing microbubbles in the vasculature in the whole brain, and performs quantitative pulsatility measurements. This novel approach can be used to obtain dynamic information along the vascular tree, in both large feeding vessels and microvessels, in-depth and in the whole brain. Such a technology could be used as an adjunct biomarker in the study of neurodegenerative diseases.

## Acknowledgment

The authors would like to thank M. Abran for technical help with the monitoring platform and M. Vanni for his advice for